\documentstyle[aps,preprint,tighten,floats,epsf,rotate]{revtex}

\begin{document}
\draft

%
\title{Skyrmion formation in 1+1 dimensions with chemical potential}
\author{V. Sunil Kumar \footnote{e-mail: sunil@iopb.res.in}, 
Biswanath Layek \footnote{e-mail: layek@iopb.res.in} , 
Ajit M. Srivastava \footnote{e-mail: ajit@iopb.res.in}}
\address{Institute of Physics, Sachivalaya Marg, Bhubaneswar 751005, 
India}
\author{Soma Sanyal \footnote{e-mail: ssanyal@indiana.edu}}
\address{Department of Physics, Indiana University, 
Bloomington 47405, USA}
\author{Vivek K. Tiwari \footnote{e-mail: vivek\_krt@hotmail.com}}
\address{Physics Department, Allahabad University, Allahabad 211002,
India}
%
%
\maketitle
\widetext
\parshape=1 0.75in 5.5in
\begin{abstract}

 Formation of topological objects during phase transitions has been
discussed extensively in literature. In all these discussions defects
and anti-defects form with equal probabilities. In contrast, many physical
situations, such as formation of baryons in relativistic heavy-ion 
collisions at present energies, flux tube formation in superconductors 
in the presence of external magnetic field, and formation of superfluid
vortices in a rotating vessel, require a mechanism which can 
bias (say) defects over anti-defects. Such a bias can crucially affect 
defect-anti-defect correlations, apart from its effects on defect density. 
In this paper we initiate an investigation for the basic mechanism
of biased formation of defects. For Skyrmions in 1+1 dimensions, we show 
that incorporation of a chemical potential term in the effective potential 
leads to a domain structure where order parameter is spatially varying. 
We show that this leads to biased formation of Skyrmions.

\end{abstract}
\vskip 0.125 in
\parshape=1 0.75in 5.5in
\narrowtext

\section{Introduction}

 Production of topological defects is of considerable importance in the 
context of the early universe. Many particle physics models 
of unification of forces predict the existence of topological defects. 
These defects may be produced during various phase transitions in the
early universe \cite{csbook}. The first detailed theory of 
formation of topological defects, in the context of the early universe,
was proposed by Kibble \cite{kbl1,kbl2}. However, the
basic physical picture of the Kibble mechanism applies equally well to the
large variety of topological defects known to arise in condensed matter
systems \cite{zrk,rjnt}. This allows the possibility of testing the
Kibble mechanism in various condensed matter systems, see refs. 
\cite{he,sccor,lc1,lc2,lc3}. In fact, the basic mechanism has many universal 
predictions which make it possible to use condensed matter experiments to 
carry out rigorous experimental tests of these predictions made for the 
defects in cosmology \cite{lc2,lc3}. The Kibble mechanism has also been 
used to study baryon formation during chiral symmetry breaking transition
in  relativistic heavy ion collisions \cite{brn1,brn2,kpst} within the 
context of Skyrmion picture of baryons. There are important issues
for the case of continuous transitions where the issue of 
relevant correlation length has been analyzed by Zurek \cite{zrk}. 
In this Kibble-Zurek mechanism, defect formation is determined by the
critical slowing down of the order parameter and depends on the
rate of transition etc. 

 In all these discussions of formation of topological defects (here, for
the sake of uniform terminology, we will use the term topological
defect even for Skyrmions, though they are not defects in the usual 
sense) defects and anti-defects form with equal probabilities. However,
there are many physical situations where formation of defects
is favored over the anti-defects (or vice-versa).  For example,
in discussing the formation of flux tubes in type II
superconductors in the presence of external magnetic field
there will be a finite density of flux tubes, all oriented along the 
direction of external field (which we call as strings, the opposite 
orientation being associated with anti-strings). In addition
there will be random formation of strings and anti-strings. Thus
the basic mechanism of defect formation here should be able to
account for the bias in the formation of strings over anti-strings.
Formation of superfluid vortices (e.g. in $^4$He) with the superfluid 
transition being carried out in a rotating vessel will also produce
biased vortex formation.

  A similar situation arises in the context of Skyrmion picture of 
baryon formation. When one wants to study baryon formation during
chiral phase transition in  relativistic heavy-ion collisions, then one
has to deal with the situation of non-zero baryon excess over antibaryons. 
This is certainly true for energies up to SPS. Even for RHIC, the baryon 
chemical potential is about 50 MeV for the central rapidity region 
\cite{rhic}. This requires that the basic
mechanism of Skyrmion formation should be able to incorporate an
intrinsic bias in favor of Skyrmions over anti-Skyrmions (depending
on the sign of the chemical potential). Such a bias can not only affect 
the estimates of net density of Skyrmions and anti-Skyrmions, it can
have crucial effects on the predictions relating to Skyrmion-anti-Skyrmion
correlations. 

 It is known that the topological models of formation of 
defects imply very specific defect-anti-defect correlations \cite{lc3}.
Predictions of these correlations, have a very important advantage over 
the predictions of defect density (as discussed in ref.\cite{lc3}).
In situations where there is no natural accurate determination of the 
correlation length,
as in the case of heavy-ion collisions, reasonable predictions of
defect (Skyrmion) density cannot be independently made. Essentially, given
a value of correlation length, any baryon density can be fitted.
In contrast, the predictions of defect-anti-defect
correlations can be put in a universal form such that they only
depend on average defect/anti-defect separation (i.e. defect/anti-defect 
density $\rho$) \cite{lc3}. 
With experimentally observed value of $\rho$, theory
of defect formation makes a unique prediction about defect-anti-defect
correlations. This is well understood for the unbiased defect/anti-defect
formation situation \cite{lc3}. Such correlations need to be understood 
for the case of biased formation, where clearly a bias of (say) defects over
anti-defects may affect these correlations in a crucial manner.  
Thus, once the theory of defect formation is extended for the 
situation of non-zero chemical potential, this can be used to make very
specific predictions about  baryon-anti-baryon correlations
in heavy ion collisions. Another important point is that in QCD the 
baryon number is strictly conserved. 
This implies that in a given region undergoing chiral phase
transition, net Skyrmion number produced must be constrained (to be
equal to the net baryon number of that region). 

In this paper we initiate a study for 
incorporating the effects of a bias in the theory of defect formation 
via the Kibble mechanism. We will first motivate the
basic ideas using physical picture of flux tube formation in the
presence of external field. We will see that this necessitates changing
the basic picture of elementary domain where the order parameter is
conventionally taken to be constant. In contrast, if the symmetry
breaking phase transition (i.e. superconducting transition) occurs 
in the presence of external magnetic field then due to the presence
of non-zero vector potential the order parameter should be taken to
vary within a domain. We mention here that the prescription of order 
parameter inside a given domain, and even more importantly, the
prescription for the inter-domain variation of order parameter (the
so called geodesic rule) is ambiguous for the case of gauge
theories \cite{gauge}. However, even apart from those considerations,
it will be clear (as discussed below) that biased formation of strings 
cannot occur unless one either modifies the geodesic rule, or introduces 
the spatial variation of the order parameter within a domain. (We mention 
here that we are assuming that the flux tube production is not dominated by
fluctuations of magnetic field, e.g. as discussed in ref.\cite{fluct}).

  We will next analyze the case of Skyrmion formation. Here a bias in
the formation of (say) Skyrmions is expected if there is non-zero chemical 
potential for baryon number. We will see that the introduction of a chemical
potential term in the effective potential for the linear chiral
sigma model will lead to spatial variation of order parameter
being preferred, having lower free energy than the case of constant
order parameter. This bias can be exactly calculated for the simpler
case of 1+1 dimensions. Even though the nature of exact bias and its
implications is complicated to analyze for higher dimensions, we 
expect similar techniques to work there. We hope to discuss that in 
a future work. 

 The paper is organized in the following manner. Section II discusses
the basic idea of biasing defects for the case of flux tube formation in a 
superconductor in the presence of external magnetic field. In rest of the 
paper we discuss the case of Skyrmions in one space dimension. In section III 
we describe the modified picture of Skyrmion formation in the presence 
of a chemical potential term.  In section IV we present the analytical 
estimate as well as numerical simulation results for the case when the 
physical space (taken for simplicity to be $S^1$, as discussed below) 
consists of only two domains. Numerical simulation results for the 3-domain 
case are presented in Sect.V. Sect. VI presents discussions and conclusions.

\section{Example of superconductor in external field}

  We first discuss the case of a type II superconductor in 2 space
dimensions. This discussion will only be qualitative. Our purpose here
is to bring out the naturalness
of spatially varying order parameter within a domain in the presence
of external magnetic field. Detailed discussion of spatial variation
for this case will be deferred for a future work as this case has
fundamental differences from the present case of Skyrmion. (For
superconductors, bias originates from non-zero vector potential
in the covariant derivative. Also 2+1 dim. case is much more
difficult to  analyze than the 1+1 dim. case which we discuss
in the present paper). 

  The free energy density for the superconductor is given by,

\begin{eqnarray}
F = |D_i \psi|^2 + a(T) |\psi|^2 + b |\psi|^4 \\
D_i \psi = (\partial_i - ie A_i)\psi ~;~ i = 1,2 . 
\end{eqnarray}

 $A_i$ are the spatial components of the vector potential and
$b > 0$, while $a(T) > 0$ for $T > T_c$ and $a(T) < 0$ for $T < T_c$. 
$T_c$ is the critical temperature.
Here, we are not including the field strength term.
We are only focusing on the nature of order parameter within a domain
in the superconducting phase from the point of view of the Kibble
mechanism. Thus full dynamics of gauge field  is not being considered
here. (As mentioned above, we are also not considering  here the 
situation when flux tube formation may arise due to fluctuations of 
the magnetic field \cite{fluct}).

 For $T < T_c$, in the superconducting state, $D_i \psi = 0$ inside
a superconductor which leads to flux quantization. In the absence of
any external fields, the formation of flux tubes during a 
superconducting transition can be studied using the Kibble-Zurek 
mechanism. The physical 2 dimensional region undergoing the phase
transition will develop a domain structure. Phase $\theta$ of $\psi$
will be constant within a domain. $\theta$ varies randomly from
one domain to another  while $\theta$ interpolates
smoothly in between two adjacent domains (see ref.\cite{gauge} for
a discussion of this point). Using this the probability of flux
vortices (for two space dimensions) can be determined to be 1/4 per 
domain \cite{prob}.

  Now consider the case when the phase transition is happening
in the presence of external magnetic field. This will require that there
be a net excess of strings over anti-strings. (Note that even with
the conventional mechanism, net string number is not zero in a given
sample. However, there the net string number could be positive or
negative, with zero average.) One way to implement
an overall excess of strings over anti-antistrings can be to impose the
net winding number at the boundary of the sample. This way one can
unambiguously fix the net excess of flux vortices over anti-vortices.
However, if one simply fixes the variation of $\theta$ at the boundary
of the sample, and keeps the conventional picture of defect formation
for the inside region (i.e. random $\theta$ for each domain and
geodesic rule for regions in-between the domains), then it is easy to
see that essentially all the excess vortices will be concentrated within 
one domain thickness of the boundary. This is because for regions which
are more than one domain distance away from the boundary, conventional
defect formation applies. This leads to defect-anti-defect symmetry 
on statistical basis for those regions. Even if there is slight excess
of vortices in the interior region in one case, anti-vortices may be
in excess in the next case (with this excess being proportional to 
$N^{1/4}$, $N$ being net number of defects+anti-defects in the interior
region). Thus, essentially entire excess vortex number imposed by the 
boundary winding will have to be carried by the one domain thick shell
within the boundary.

 Similar situation can be argued even for the case
of Skyrmions. It is important to realize that net winding for integer
winding Skyrmions cannot be fixed in this manner by prescribing a 
winding at the boundary. This is because each full integer winding 
Skyrmion represents a compactified region of space where the boundary of
a ball (disk in 2 space dimensions and 3-ball in 3 space dimensions) has
a constant value of the chiral field. Thus there is no memory of winding
of Skyrmion at the boundary of the region. However, as discussed in the
literature \cite{brn2}, integer winding Skyrmions almost never form during
a phase transition. It is only the partial winding Skyrmions which have
reasonable probability of formation. The winding numbers of these 
partial winding Skyrmions can be fixed by prescribing the winding at
the boundary. For example, for 2-d Skyrmions with Skyrmion winding 
being carried by the polar angle $\theta$ and the azimuthal angle 
$\phi$ on $S^2$, one can assume that $\theta$ 
never approaches $\pi$ outside the partial Skyrmions (with $\theta = 0$
at the center of the Skyrmion). Then $\phi$ windings will determine
whether one has Skyrmion or anti-Skyrmion. Fixing $\phi$ winding at the
boundary of the 2-dimensional region will fix the net Skyrmion
number, assuming that these partial Skyrmions will evolve to become
integer Skyrmions. (Actual situation is more complicated as the
variation of the polar angle $\theta$ may in general be different
in different partial Skyrmion regions.) It may thus be possible
that one will end up with the same situation as
discussed above for the case of superconductor. That is, if a net
Skyrmion winding is fixed at the boundary of the region, then the excess
Skyrmions will all be essentially concentrated within few domains thick
shell at the boundary.

  Clearly this situation is not satisfactory. When excess flux vortices 
form due to external field, they are roughly uniformly distributed in the
sample. Similarly, non-zero baryonic chemical potential will lead to
a uniform excess in the baryon density over anti-baryon density. It is then 
clear that one needs to modify the whole picture of defect formation itself.
As we will discuss below, fixing boundary windings can be used to implement
exact conservation of defect number, say Skyrmion number. (This exact
conservation may not be needed for the superconductor case where a
net statistical excess of strings may be sufficient.) However, this
has to be supplemented with modification in the basic defect formation
picture in order to distribute the excess defects uniformly in the whole 
region. We now discuss this modification of the defect formation picture.

 Let us again consider the case of a 2-dimensional region undergoing 
superconducting transition in the presence of external field. Until the 
order parameter field $\psi$ settles to its thermodynamic equilibrium value, 
there will be non-zero currents present and $D_i \psi$ will not necessarily
vanish inside the superconductor. As $\psi$ approaches the equilibrium
value, the magnetic field will be pushed out of the superconducting
regions, getting trapped in the form of isolated vortices, and $D_i \psi$
will approach zero inside the superconducting region. If $\theta$ is
taken to be constant inside each domain, as in the case of zero 
external field, then $A_i$ will also be constant inside each domain.
This will require modification for the standard geodesic rule to account
for extra vortices due to external field. Though this possibility cannot
be easily ruled out (and needs to be examined carefully), it is
certainly not very natural. This leads to the requirement that the biased
formation of vortices over anti-vortices is entirely due to regions
in-between the domains. Certainly, for a second order transition, the
splitting of the region in terms of elementary domains and inter-domain
regions is somewhat ambiguous. Domains do not have completely
constant $\theta$ with all the $\theta$ variation remaining in the
inter-domain regions. In fact, one can very well consider the
inter-domain regions as elementary domains with original domains 
becoming the inter-domain regions. Statistically it should not make
much difference in calculation of defect formation probabilities. 
Thus, if vortices are to be biased over anti-vortices, then it looks
unnatural to assume that the domain picture remains unchanged while
{\it only} the inter-domain regions implement the biased vortex formation. 

 A more satisfactory picture, which as we will see below, also naturally
fits with the picture of Skyrmion formation with non-zero chemical
potential, is the following. As there is
a net magnetic field going through the region, we assume that $\theta$
inside each domain has spatial variation. With $D_i \psi = 0$ inside
each domain, it will imply specific biased variation of the vector
potential inside each domain. Combined with a suitably modified geodesic
rule, this will lead to an overall excess of flux vortices over
anti-vortices. (Similarly, if we consider superfluid vortices 
forming during phase transition in a rotating vessel, then due to already 
present fluid velocity, a non-zero gradient of the condensate phase should 
naturally arise.) For the superconductor case, the
exact nature of the spatial variation of $\theta$
(and consequently of $A_i$) will depend on the choice of gauge etc. 

  There are several points here that one should be careful about. First,
as mentioned above, the dynamics of flux tube formation may be dominated 
by fluctuations of magnetic field, as discussed in ref.\cite{fluct}. 
The considerations 
discussed here do not apply to such cases. Second point is that the uniform
field at the beginning of phase transition will become unstable at the
end of the transition, after all that is how the flux tubes form. The choice
of vector potential to bias $\theta$ variation should take into account
such a dynamics of the magnetic field. These are complex issues and need
to be examined in detail. These issues for the case of 2-dimensional 
superconductor, and for superfluids, will be discussed in a future work.
The arguments we have discussed for this case
are very crude. Our purpose here was only to physically motivate that
spatially varying order parameter within a domain can arise in a physical
example. As we will see below, same effect can arise by postulating a
chemical potential term for the free energy for the Skyrmion case.

\section{Skyrmions in 1+1 dimensions}

 Above, we have tried to argue qualitatively what kind of modifications
are needed for the basic domain picture to implement the biased formation 
of defects. We now consider the case of Skyrmion formation in one space 
dimension where these ideas can be concretely implemented. 
To represent the situation of a nonzero chemical potential, we will
add a $-\mu N$ term to the effective potential of linear chiral
sigma model (see ref.\cite{mu} in this context). Here $N$ is the 
topological Skyrmion winding number to be identified with the baryon number
(in 3+1 dimensions). In one space dimension we can write the appropriate free
energy (for possibly inhomogeneous field configurations) as follows. For
simplicity of presentation we use dimensionless variables. We divide
the free energy density by $m^2$, the coefficient of $\phi^2$ term. 
Length scales are measured in units of $m^{-1}$, and the chemical 
potential $\mu$ in units of $m$. We can now write down the 
dimensionless free energy as,

\begin{eqnarray}
F = \int [|\partial_x \phi|^2 - |\phi|^2 + 
\lambda |\phi|^4 - \mu n(x) ] dx, \\
n(x) = {1 \over 2\pi} \partial_x \theta.
\end{eqnarray}

 Here $N = \int n(x) dx$ is the Skyrmion number. $\phi$ takes 
values on a circle $S^1$, which can be written as $\phi = \phi_0 
e^{i \theta}$.(We mention that there is no strictly conserved topological 
current for the case of linear sigma model, hence no exact conservation of
Skyrmion number. This is why, to implement exact conservation of baryons
we will need to fix the winding number at the boundary of the sample.)

 Let us determine what configuration of $\phi$ is preferred for the 
domain. In one space dimension only topological configurations are the 
Skyrmions. For these, the magnitude of $\phi$ can be 
taken to be constant $\phi_0$ for the entire 
1 dimensional physical space. The quadratic and quartic terms in $\phi$
in Eq.(3) then add up to give a constant which we can neglect in the
following discussion. The relevant terms in $F$ from Eq.(3) are,

\begin{equation}
F_d  = \int [\phi_0^2 (\partial_x \theta)^2 - {\mu \over 2\pi} 
\partial_x \theta ] dx.
\end{equation}

 Note that, Skyrmion number $N$ is a topological invariant only when
boundary conditions are fixed. Further, we determine the local value of 
equilibrium order parameter by minimizing the free energy {\it density}
hence $n(x)$ can affect the determination of the equilibrium value
of the order parameter as we show below.
The chemical potential term, being a total derivative, does not
contribute to the field equation $\partial_x^2 \theta = 0$. A general 
solution of this equation is $\theta(x) = \alpha x + \theta_0$.
As the free energy density in the expression
for $F_d$ depends only on $\partial_x \theta = \alpha$, we minimize 
it with respect to $\alpha$ to get,

\begin{equation}
\alpha = \partial_x \theta = {\mu \over 4 \pi \phi_0^2} ~ 
\equiv ~ \mu^\prime.
\end{equation}

 We therefore see that for non-zero $\mu$, constant $\theta$ is not
preferred as an equilibrium configuration. This is an unconventional
result. Normally, for a theory where some internal symmetry is
spontaneously broken, the vacuum is translationally invariant.
This is how the effective potential is calculated, starting from 
the effective action, and restricting to constant field configuration.
Spatial variations of the order parameter are conventionally discussed 
only as departure from the equilibrium situation. The non-trivial part of
the above equation is that spatial variation of the order parameter
is present even for the equilibrium configuration of the order parameter.
This is happening due to the fact that the chemical potential term 
for the Skyrmion number is expressed in terms of the derivative 
of the order parameter field $\phi$.
As long as derivative terms come with positive definite contributions
to the free energy, the equilibrium order parameter configuration
(i.e. with lowest free energy) will be spatially constant. This is what
happens with the conventional $(\partial_i \phi)^2$ term. Even the
presence of Skyrme term \cite{skrm} will lead to the same conclusion,
i.e. constant order parameter.
The situation is completely different with the above chemical potential
term. This term is linear in the derivative of $\phi$ and hence not
positive definite. Therefore, it is no longer certain that the
equilibrium configuration will be that of a constant $\phi$. It will
be interesting to investigate how one can justify this type of picture at
a more fundamental level. For example, for QCD with fundamental degrees
of freedom being quarks and gluons, a chemical potential
term for the baryon number apparently has no reason to lead to a
situation where the equilibrium order parameter should have spatial 
variations. However, as we have argued above, in the chiral models,
the chemical potential term for the baryon (Skyrmion) number should
naturally lead to spatial variation for the equilibrium order parameter.  
In this context, we mention ref.\cite{mu} where a chemical potential
term for the Skyrmion number has been derived.

  With Eq.(6), our strategy for determining the formation
of Skyrmion will be as follows. We split the physical space, say x axis, 
in terms of elementary domains which are one dimensional segments for the
present case. We choose random values of $\theta$ in the middle of
each domain. In a given domain, the spatial variation of $\theta$ from one
end to the other end is then determined by integrating Eq.(6). 
With uniform variation of $\theta$ across the domain, net $\theta$ 
variation within a given domain of size equal to correlation length $\xi$, 
will be,

\begin{equation}
\Delta \theta = {\mu \xi \over 4 \pi \phi_0^2} ~=~ \mu^\prime \xi.
\end{equation}  
 
  We still need to specify how $\theta$ is supposed to interpolate 
in between two adjacent domains. Conventionally (i.e. with $\mu = 0$)
one simply uses the geodesic rule which is the statement of minimizing
the free energy in the inter-domain region. Note that this geodesic
rule does not require specification of how large the inter-domain region 
actually is. With non-zero $\mu$, one needs to specify more details about
the inter-domain region. We will still follow the spirit of the geodesic
rule, i.e. minimize the net free energy in the inter-domain region. However,
due to the chemical potential term it does not lead to the conventional
geodesic rule. Depending on the value $\theta$ at the end of one domain,
and the value $\theta^\prime$ at the beginning of the next domain, it 
may happen that the longer path on the vacuum manifold $S^1$ may be 
preferred in some cases. An example of this is shown in Fig.1 where, due to
the chemical potential term, the anti-clockwise variation from $\theta$ to 
$\theta^\prime$ may have lower free energy compared to the shorter clock-wise
variation. We compare the values of $F_d$ (Eq.(5)) for the two paths
on $S^1$, with opposite orientations, joining $\theta$ and $\theta^\prime$.
The path with lower value of $F_d$ is selected to represent the $\theta$
variation in the inter-domain region. We take uniform variation of $\theta$
for both paths, any other variation should give higher free energy for
a given orientation of path (unless the chemical potential term in Eq.(5)
becomes completely dominant). The choice of the path not only depends on 
$\theta$ and $\theta^\prime$, but also on the domain size $\xi$ as well as
the size of the inter-domain region (and obviously on $\Delta \theta$ in
Eq.(7)). We therefore introduce another length scale $\chi$ which represents 
the size of the inter-domain region. We will present results for varying 
values of $\xi$ and $\chi$. These two different length scales have a clear
meaning in a first order transition via bubble nucleation. The core
of bubble representing uniform magnitude of the order parameter will
have size $\xi$ while the bubble wall region for two coalescing bubbles
can represent the inter-domain region with size $\chi$. (Just like
the conventional case of $\mu = 0$, free energy in the inter-domain
region will be higher which, for the bubble case, can come from the energy
of coalescing bubble walls). For a second order transition, definition
of domain and inter-domain regions is much more fluid. One can take
$\xi$ to represent the region where the order parameter variation is given by
the field equation (Eq.(7)) whereas, in the inter-domain region larger
variations of $\theta$ may occur.

  Using Eqs.(5) and (7), it is easy to show that, for $0 < \theta < 
\theta^\prime < 2\pi$, the anti-clockwise variation from $\theta$
to $\theta^\prime$  has lower free energy compared to the clockwise
variation if the following condition is satisfied,

\begin{equation}
\theta^\prime - \theta < \Delta \theta {\chi \over \xi} + \pi .
\end{equation}

  Here $\theta$ and $\theta^\prime$ are the values of the phase at
the two end points of a given inter-domain region (i.e., say, $\theta$
at the end of one domain and $\theta^\prime$ at the beginning of the
next domain). From Eq.(8) we see directly that for $\Delta \theta > 0$,
anti-clockwise variation may be preferred even when this variation
is larger than $\pi$. 

  Due to this detailed rule for determination of $\theta$  
in the inter-domain region, the final probability becomes dependent on the 
length scales $\xi$ and $\chi$ in a complicated manner. It then 
becomes very difficult to determine analytically the probability of
Skyrmion formation in this case. In the following, we will present an
analytical calculation of the probability for the two domain case, but
for three domain case we will need to resort to numerical
simulation for determination of this probability. 

 Another important point is that the formation of Skyrmion is actually
very different from the formation of other topological defects like
monopoles, strings etc. This is because Skyrmions require constant
boundary conditions for their topological description. These issues
have been discussed in the literature. One determines the
probability of formation of partial winding Skyrmion which is
expected to subsequently evolve into a full integer winding 
Skyrmion \cite{brn2}. However, these complications are not relevant for
the main point of the present work which is to investigate the effects of
non-zero chemical potential on the basic theory of defect formation. 

 To avoid the issues of boundary conditions etc. for Skyrmion case,
we will further simplify the problem. For the present case of one space
dimension, we consider the physical space to
be compact, i.e. a circle $S^1$. Further we consider the size of the 
circle to be sufficiently small so that it can accommodate only a couple of
correlation domains which are sufficient to form one Skyrmion
in the whole circle. This way the issue of boundary condition becomes
irrelevant. For $\mu = 0$ case this would have been the same as the
calculation of the probability of winding around a loop going through
couple of domains, i.e. the probability of vortices. 

\section{two domain case}

  We know that for $\mu = 0$ case there is no possibility of any
winding unless one considers three or more domains. This is not the
case for $\mu \ne 0$ case. Due to the spread of $\theta$ (given by
Eq.(7)) within a given domain, it is easy to see that even if the physical
space $S^1$ has only two domains, one can still get non-trivial windings.
Each domain already contains partial positive windings (for $\mu > 0$).
Inter-domain regions only need to complete this winding. It is also
easy to convince oneself that there is no possibility of getting any 
anti-winding with only two domains (again, for $\mu > 0$ case). 
[Positive windings correspond to the case when, as one goes 
around the closed path in physical space in an anti-clockwise/clockwise
direction, then the variation of $\theta$ in the order parameter space $S^1$
is also in an anti-clockwise/clockwise direction. Negative windings
will correspond to the case when the direction of variation of $\theta$
in the order parameter space is opposite to the direction of the path
in the physical space.]

  Consider the physical space $S^1$ (as discussed above) consisting
of two domains as shown in Fig.2. The two random values of $\theta$ at the 
centers of the two domains are $\alpha$ and $\beta$. The anti-clockwise 
variation of $\theta$ as given by Eq.(7) within each domain is shown in the
figure. Arrows in the figure show anti-clockwise path taken in the physical 
space to calculate the winding of $\theta$. For notational simplicity, we 
denote the beginning values of $\theta$ (along the anti-clockwise path in 
the physical space) in each of the two domains as follows,

\begin{equation}
\alpha - {\Delta \theta \over 2} \rightarrow \theta_1 , \\
{\rm \quad and \quad }
\beta - {\Delta \theta \over 2} \rightarrow \theta_2. \\
\end{equation}

With this, the values of $\theta$ at the other ends of each domain
become $\theta_1 + \Delta \theta$ and $\theta_2 + \Delta \theta$
respectively. In Fig.2, domains (of size $\xi$) are 
shown by the dashed curves and inter-domain regions (of size $\chi$) are 
shown by the dotted curves. 

 We will set $\theta_1 = 0$, so that all other angles shown in Fig.2
are less than 2$\pi$ and are measured from $\theta_1 = 0$. We also assume 
that $\Delta \theta$ as given in Eq.(7) is less than $\pi$ to simplify
the analytical estimates.  Within each
of the two domains the variation of $\theta$ is equal to $\Delta \theta$
(as given in Eq.(7)) and is in the anti-clockwise direction on the
vacuum manifold $S^1$. The formation of defect or antidefect then depends
on whether in the two inter-domain regions $\theta$ interpolates in
the clockwise direction or in the anti-clockwise direction. 
(We again repeat that we are using the terms defect and antidefects
even for Skyrmions and anti-Skyrmions, though they are not defects in
the usual sense). Let us call the inter-domain region from $\theta_1 + 
\Delta \theta \rightarrow \theta_2$ as Junction 1 and the inter-domain 
region from $\theta_2 + \Delta \theta \rightarrow \theta_1$ as Junction 2. 
Then there will be four cases depending on the direction of the 
variation of $\theta$ at these two junctions.
 
(1): Junction 1 clockwise, Junction 2 anti-clockwise

(2): Junction 1 anti-clockwise, Junction 2 clockwise

(3): Junction 1 clockwise, Junction 2 clockwise

(4): Junction 1 anti-clockwise, Junction 2 anti-clockwise

 As $\theta_1$ has been set equal to zero, the value of $\theta_2$ will
decide which of the above cases is preferred as the lowest free energy
configuration, and hence the resulting winding number.

First consider the situation when, on the vacuum manifold $S^1$, 
$\theta_2$ lies in between $\theta_1 (= 0)$
and $\theta_1 + \Delta \theta (= \Delta \theta)$. It is easy to see
that in this situation, cases (1) and (2) give rise to one positive
winding each, case (3) gives no winding and case (4) leads to a 
positive winding number 2 defect.  

 Using Eq.(8), we see that for case (4) the requirement of
anti-clockwise variation at Junction 1 implies

\begin{equation}
2\pi - \Delta \theta + \theta_2 < \Delta \theta \frac{\chi}{\xi} + \pi
\quad \Rightarrow \quad
\theta_2 < \Delta \theta(\frac{\chi}{\xi} + 1) -\pi .
\end{equation}

 If we assume that the values of chemical potential, the domain
size $\xi$, and the inter-domain size $\chi$ are such that,

\begin{equation}
\Delta \theta < {\pi \over (\frac{\chi}{\xi} + 1)} ,
\end{equation}

\noindent then we see that the requirement at junction 1 cannot be satisfied 
since $\theta_2 > 0$. Thus no winding 2 defects will form in this
case. We will assume this to be the case for the analytical estimates.
Thus, here, we are calculating the probability of defects when the bias 
given by the chemical potential (Eq.(7))
is not very large. We do this for the sake of simplicity as analytical
estimates for large $\Delta \theta$ become more complicated (with larger
variety of configurations contributing to defect formation). When we 
present numerical results, then we will present results for larger
values of $\Delta \theta$ also.

 For case (2) we see that the requirement on Junction 1 is the
same as in case (4). With the requirement on $\Delta \theta$
as given in Eq.(11), we thus see that no defects will form for the case (2).

 Now let us discuss case (1). The conditions on $\theta_2$ from Junction 2 
and Junction 1 are, respectively,

\begin{eqnarray}
\theta_2 > - \Delta \theta (\frac{\chi}{\xi}+1) + \pi \equiv A \\ 
\theta_2 >  \Delta \theta (\frac{\chi}{\xi}+1) - \pi = - A .
\end{eqnarray}

 Note that with Eq.(11), $A > 0$. Also, we are discussing the case when
$0 < \theta_2 < \Delta \theta$ hence the allowed range for $\theta_2$ 
(from Eq.(12)) is $\Delta \theta - A = \Delta \theta (\frac{\chi}{\xi}+2) 
- \pi$. Thus, if we make a stronger restriction (compared to that given by
Eq.(11)) on $\Delta \theta$ such that,

\begin{equation}
\Delta \theta < {\pi \over (\frac{\chi}{\xi} + 2)} ,
\end{equation}

\noindent  then no defects will form in case (1) also. We will assume 
that Eq.(14) is satisfied by the parameters so that no defects are formed
when $\theta_2$ lies between $0 (= \theta_1)$ and $\Delta \theta$.
  
  Let us now consider the situation when $\theta_2 > 2\pi - \Delta \theta$.
(Note, $\theta_2 < 2\pi$.)
Again completing path anticlockwise in the physical space in Fig.2, one
can see that in this situation also, cases (1) and (2) give rise to one 
positive winding (assumed to be anti-clockwise winding) each, case (3) 
gives no winding and case (4) leads to a winding number 2 defect. Following
the arguments as above for all the four junction conditions, one can
show that, with Eq.(14), no defects are formed in this case. 

 Finally we discuss the situation when $\theta_2$ takes any value in the rest
of the vacuum manifold, i.e. $\Delta \theta < \theta_2 < 2\pi - 
\Delta \theta$. In this situation one can see that cases (1) and 
(2) do not give any defects. Case (3) gives an anti-defect
whereas case (4) gives formation of winding one defect. As we have
mentioned earlier, for the two domain case no anti-winding defects can
form. One can see it here directly also by requiring the conditions on 
Junction 2 and Junction 1 for case (3), i.e.

\begin{eqnarray}
2\pi - \theta_2 - \Delta \theta > \Delta \theta \frac{\chi}{\xi} + \pi \\
\theta_2 - \Delta \theta > \Delta \theta \frac{\chi}{\xi} + \pi ,
\end{eqnarray}

respectively. These give,

\begin{eqnarray}
\theta_2 < - \Delta \theta (\frac{\chi}{\xi} + 1) + \pi \\
\theta_2 > \Delta \theta (\frac{\chi}{\xi} + 1) + \pi .
\end{eqnarray}

These conditions are mutually inconsistent since  $\Delta \theta 
(\frac{\chi}{\xi} + 1) > 0$. Thus no antidefects form.

  We thus consider the last case (4). Junction 2 condition gives

\begin{equation}
\theta_2 > -\Delta \theta (\frac{\chi}{\xi} +1 ) + \pi .
\end{equation}

Junction 1 condition gives,

\begin{equation}
\theta_2 < \Delta \theta (\frac{\chi}{\xi} +1 ) + \pi .
\end{equation}

 One can easily check that these conditions are consistent with the range
of $\theta_2$ for this case, i.e. $\Delta \theta < \theta_2 < 2\pi - \Delta 
\theta$.

   Thus, the allowed range of $\theta_2$ for the formation of winding
one defects is 

\begin{equation}
\theta_2^{\rm Range} = (2\Delta \theta) (\frac{\chi}{\xi} +1 ) .
\end{equation}

 As $\theta_2$ can take any value between $0$ and $2\pi$, we conclude that
the probability for the formation of winding one defects for the two domain
case is

\begin{equation}
P = \frac{(\Delta \theta)(\frac{\chi}{\xi} +1)}{\pi} .
\end{equation}
 
 This expression is valid for the values of parameters such that Eq.(14)
is satisfied. No antiwinding defects are formed for the two domain case.

  We have carried out numerical simulation for the two domain case by 
randomly assigning values of $\theta_1$ and $\theta_2$ (note, that this is 
completely equivalent to assigning random values of $\theta$ at the middle of 
the domains) as shown in Fig.2. For each of the two junctions, 
we then compare the free energies (Eq.(5)) for the clockwise
and the anti-clockwise variations of $\theta$ around the vacuum manifold
and select the path with lower free energy. Net winding is then calculated
for the anti-clockwise path in the physical space starting from,
and ending at $\theta_1$. Fig.3 shows the results of numerical simulation.
Symbols give the numerical simulation values of the probability $P$ 
which is the probability of positive windings as a function of $\Delta 
\theta$. (Here, as $\Delta \theta$ is varied, $\xi$ is kept fixed
in Eq.(7).)  Solid continuous lines show the plots of the analytical 
results in Eq.(22). This is plotted only up to the value of $\Delta \theta = 
\frac{\pi}{(\frac{\chi}{\xi} +2)}$ as allowed by Eq.(14).  We see that 
simulation results completely agree with the analytical results. We had 
carried out simulation for 100,000 events so the numerical errors are 
negligible.  One point here is that, although Eq.(22) seems to imply that 
only the ratio $\chi/\xi$ is relevant for the probability calculation, it is
because we have restricted to small $\Delta \theta$ case. For larger
values of chemical potential, the probability depends more generally 
on $\xi$ and $\chi$, though it becomes complicated to work out all 
possible configurations. We also note that for $\xi = 1, \chi = 2$,
probability of 2-winding Skyrmions also becomes non-zero beyond
$\Delta \theta \simeq 1.1$. For values of $\Delta \theta$ larger than
this, the total probability plot for this case also includes the
contributions from these 2-winding Skyrmions. (We do not plot the
probabilities beyond $\Delta \theta \simeq \pi/2$ as for higher values
even higher winding Skyrmions start forming and correspondence with
the analytical estimates becomes more remote.) Double winding Skyrmions
are counted as 2 times single winding Skyrmion. In this sense 
the total probability for this case represents the average number 
of positive winding Skyrmions. 

\section{three domain case}

  For three domain case we do not attempt analytical estimate as it becomes
very complicated. Physical space is divided in three domains (with three
inter-domain regions) in a similar manner as the two domain case in Fig.2.
Random values of the three angles in the three domains are chosen
(at the beginning of each domain, which, as discussed for the two
domain case, is equivalent to selecting random values at the middle
of each domain). The path on the vacuum manifold is numerically determined 
for each inter-domain region by calculating free energies as before.

Figs.4 and 5 show plots for the case of three domains. Fig.4a shows the
plots of the total probability $p_+$ of Skyrmion formation (giving average 
positive winding Skyrmion). We see that this increases with increasing 
value of $\Delta \theta$ (and hence $\mu$). 
Total probability is  calculated by multiplying
the number of Skyrmions of a given winding number by the winding number.
The value of this probability $p_+$ thus refers to average positive
winding Skyrmion number. Fig.4b shows the  probability $p_-$ of anti-Skyrmion 
formation. AntiSkyrmions are found only with winding minus one for this
three domain case. In Fig.4, we can 
make comparison with the conventional case of $\mu = 0$. We thus see that
plots of winding plus one and minus one intersect the y axis ($\mu = 0$) 
at $p_+ \simeq p_- \simeq 0.125$ (corresponding to the net probability = 
1/4 of defect or anti-defect formation). We see that non-zero $\mu$ affects 
the probabilities very strongly.  Interesting thing to note is that the 
net probability of formation of Skyrmion or anti-Skyrmion also increases 
with $\Delta \theta$, as shown in Fig.4c.

  Fig.5a-d show plots of probabilities (again, average Skyrmion number)
for positive winding Skyrmions for windings +1, +2, +3, and +4 respectively.
(Note that for the dotted curve in Fig.5a, the probability of winding one
Skyrmion becomes almost zero beyond $\Delta \theta \simeq 1.5$, which
is responsible for the change of slope for the corresponding plot in
Fig.4a.) We plot probabilities for values of $\Delta \theta < \pi$ (for 
Fig.4 and 5, except Fig.4b where plot is given only upto $\Delta \theta$ 
= 0.75 as $P_-$ becomes zero beyond this value).  This is because of two 
reasons. First it already illustrates the basic features 
of probabilities of defect and antidefect formation in the presence
of a bias. Secondly, already in this case very complex Skyrmion
configurations (with winding numbers 3 and 4) start forming. The
plots of total probability are then calculated by multiplying
the number of Skyrmions of a given winding number by the winding number.
The value of the probability thus refers to average Skyrmion number.
For values of $\Delta \theta$ larger than $\pi$ even larger winding
Skyrmions (winding 5) start forming and it becomes difficult to 
extract any pattern from the plots. In this sense the parts of the
plots for smaller values of $\Delta \theta$ ($< \pi/2$) are most
physical as they primarily contain Skyrmions up to winding 2, and one can
get a feeling for how probabilities are changing with changing
$\Delta \theta$.

  Another point is that the probabilities depend in a complex manner
on the two length scales $\xi$ and $\chi$. Though one may expect that
$\theta$ variation within a domain, and its bias in the inter-domain 
region both should have similar effects. But our plots do not imply that.
In fact, there is an important difference between the region of the domain
and the inter-domain region in our model. Inside a domain, variation of 
$\theta$ is fixed by Eq.(6). Whereas in the inter-domain region the variation
is fixed by the boundary conditions at the ends of the region and
crucially depends on the length scales $\xi$ and $\chi$. It may be 
interesting to investigate this issue further, possibly by developing 
some scheme which treats the domain and inter-domain regions on 
a somewhat equal footing.

  We have thus achieved our goal, that is, we have 
developed a mechanism which uniformly
biases the formation of windings of one sign over the opposite windings.
It is very satisfying that a very natural extension of the sigma 
model effective potential with a chemical potential term leads to
qualitatively similar picture of spatially varying $\theta$ within
an elementary domain as one expected in the case of flux tube formation 
for superconducting transition in the presence of external field (or
the superfluid case).

\section{Conclusions and Discussions}

 We mention an interesting possibility which arises from the observation
in Fig.4
that anti-Skyrmion probability $p_-$ drops to zero at some value of 
$\Delta \theta$ (corresponding to some specific value of $\mu^\prime$) 
before the Skyrmion probability (average Skyrmion number) $p_+$ crosses 
the value 1. If one can get a similar plot for the realistic 3+1 
dimensional case, then the value of chemical potential $\mu_-$, beyond which 
$p_-$ drops to zero, may have some relationship with the critical 
value of the chemical potential $\mu_c$ in the QCD phase diagram at 
T = 0 (for massless flavors, as explicit symmetry breaking has not been 
accounted for in the present model). This is because, for $\mu > \mu_-$ 
in our model, the effects of phase transition induced domain-to-domain
fluctuations, (thermal or quantum), become sub-dominant compared
to the effects of the bias generated by $\mu$, thereby leading to zero 
anti-baryon production.  Any further increase in $\mu$ only creates more 
baryons, consistent with increasing $\mu$, without creating any anti-baryons. 
It seems reasonable to expect that for $\mu < \mu_c$ in the QCD phase diagram, 
the chiral transition (by varying temperature)  should be able to produce at 
least some anti-baryons also, especially due to non-equilibrium nature of the 
domain structure of the chiral order parameter. For $\mu > \mu_c$ there is
no phase transition, and for $T = 0$ one does not expect any anti-baryons 
to be present. Similar thing is happening here where no anti-baryons
are produced in our model for $\mu > \mu_-$, suggesting a possible 
relationship of $\mu_-$ with $\mu_c$. One could also explore whether
$\mu_c$ could be related to the value of $\mu$ in our model beyond which
the Skyrmion probability (average Skyrmion number) crosses 1. However, 
in that regime every domain will have a Skyrmion on the average and 
the situation will resemble less like the situation of a phase transition.
 
 We note here that a term, similar to the chemical potential term in Eq.(3), 
has been used previously in the context of certain models of electroweak 
baryogenesis where, to favor baryons over anti-baryons, a CP violating term 
is introduced by writing an effective chemical potential term for the
Chern-Simons number \cite{turk}. However, there this term corresponds
to the winding number of the gauge fields, and not the Higgs field as
in the present case. Hence, the chemical potential term in ref.
\cite{turk} does not affect the vacuum configuration. It affects
the evolution of the fields, thereby biasing the dynamics in favor of
positive change in the Chern-Simon number. If a term like we use in the
present paper can be written down for the Higgs field also in those 
models \cite{turk}, then our results of the present paper suggest that 
one may be able to produce baryon asymmetry using biased production of 
texture configurations over anti-texture configurations (apart from any 
bias coming from dynamics). At the same
time, discussions of the CP violating term in ref.\cite{turk} suggest 
that one may be able to bias production of Skyrmions over anti-Skyrmions
by biasing the field evolution dynamics in favor
of Skyrmions over anti-Skyrmions. Though here it would happen in different
way compared to the case in ref.\cite{turk} where competing time scales of
the Higgs field evolution and gauge field evolution play a crucial role.
In the context of chiral models, one will utilize the fact that, as 
mentioned above, initially only partial winding Skyrmions are expected to 
form which evolve to full winding Skyrmions subsequently. If the dynamics
could be favored for positive windings compared to negative windings then
one can imagine that a large number of initial partial positive winding 
configurations will have chance of evolving to (positive) integer winding,
while a lesser number of configurations will be able to evolve to (negative)
integer windings. It will be interesting to investigate these
possibilities.

 Many steps remain to complete the program initiated in the present work. 
First, one needs to
investigate the formation of Skyrmions in a one-dimensional space of large
extent where issues of boundary conditions will play a role.
Though, non-trivial aspects of evolution of partial winding Skyrmions
into full winding Skyrmions are expected to occur only in 2 or higher
3 space dimensions. For 2 and 3 dimensions one needs to work out
the exact nature of spatial variation of the chiral order parameter
within a domain as well as for inter-domain regions. Then the above 
described program will be easily extended for these dimensions as well.
Also, in the context of chiral phase transition in QCD, one would like 
to implement exact Skyrmion (baryon) number conservation. That is, for a 
given volume of physical space, defect formation mechanism should be able to
produce a given amount of net Skyrmion excess over anti-Skyrmion.
This is to represent the experimental situation where a given 
volume of quark-gluon plasma (QGP) in chirally symmetric state undergoes 
chiral symmetry breaking. Resulting net baryon number should exactly equal 
the baryon number contained in the original QGP region. (Neglecting
any baryon diffusion across the boundary of the region under 
consideration.) As mentioned above, we can now achieve it in the 
following manner. We fix the winding at the boundary of the whole region  
to exactly equal to the required baryon number inside the boundary (i.e.
the initial baryon number of the QGP volume). The problem of concentration
of excess baryons near the shell (as mentioned above) can now be avoided
by using the modified defect formation method for the interior region. 
The value of $\mu$ can be adjusted suitably such that it leads to
uniform excess of baryon density from the shell to the interior region.

  In conclusion, we have proposed a method to handle production of
topological defects when physical situation requires excess of 
windings of one sign over the opposite ones. We have considered the case
of Skyrmions in 1+1 dimensions and have shown that addition of a
chemical potential term leads to modification in the order parameter 
distribution inside elementary domains leading to excess production of
Skyrmions over anti-Skyrmions (for $\mu > 0$, for negative $\mu$ 
anti-Skyrmions are preferred over Skyrmions). Extension of these techniques
to the 3 dimensional case will give us a realistic Skyrmion
formation mechanism which can be applied for the case of relativistic
heavy-ion collisions. Same ideas will also help to investigate formation
of flux vortices in superconductors in the presence of external magnetic
field, as well as formation of superfluid vortices when the transition
is carried out in a rotating vessel. As our results show, biased formation 
of defects can strongly affect the estimates of net defect density. Also, 
these studies may be crucial in discussing the predictions relating 
to defect-anti-defect correlations.  Once the theory of defect formation 
is extended for the situation of non-zero chemical potential, this can 
be used to make very specific predictions about  baryon-anti-baryon 
correlations in heavy ion collisions.

\vskip .2in
\centerline {\bf ACKNOWLEDGMENTS}
\vskip .1in

  We are very thankful to A.P. Balachandran and Balram Rai for many
useful discussions and comments. We also thank Sanatan Digal, Hari Kumar, 
Ananta P. Mishra, Rajarshi Ray and Supratim Sengupta for useful 
suggestions. AMS and VKT acknowledge the support of the Department
of Atomic Energy- Board of Research in Nuclear Sciences (DAE-BRNS),
India, under the research grant no 2003/37/15/BRNS/66. SS would like 
to thank the Nuclear Theory Center, Indiana University, Bloomington for
hospitality. VKT would also like to thank the Physics Department, D.D.U.
Gorakhpur University for support.



\begin{figure}[h]
\begin{center}
\leavevmode
\epsfysize=20truecm \vbox{\epsfbox{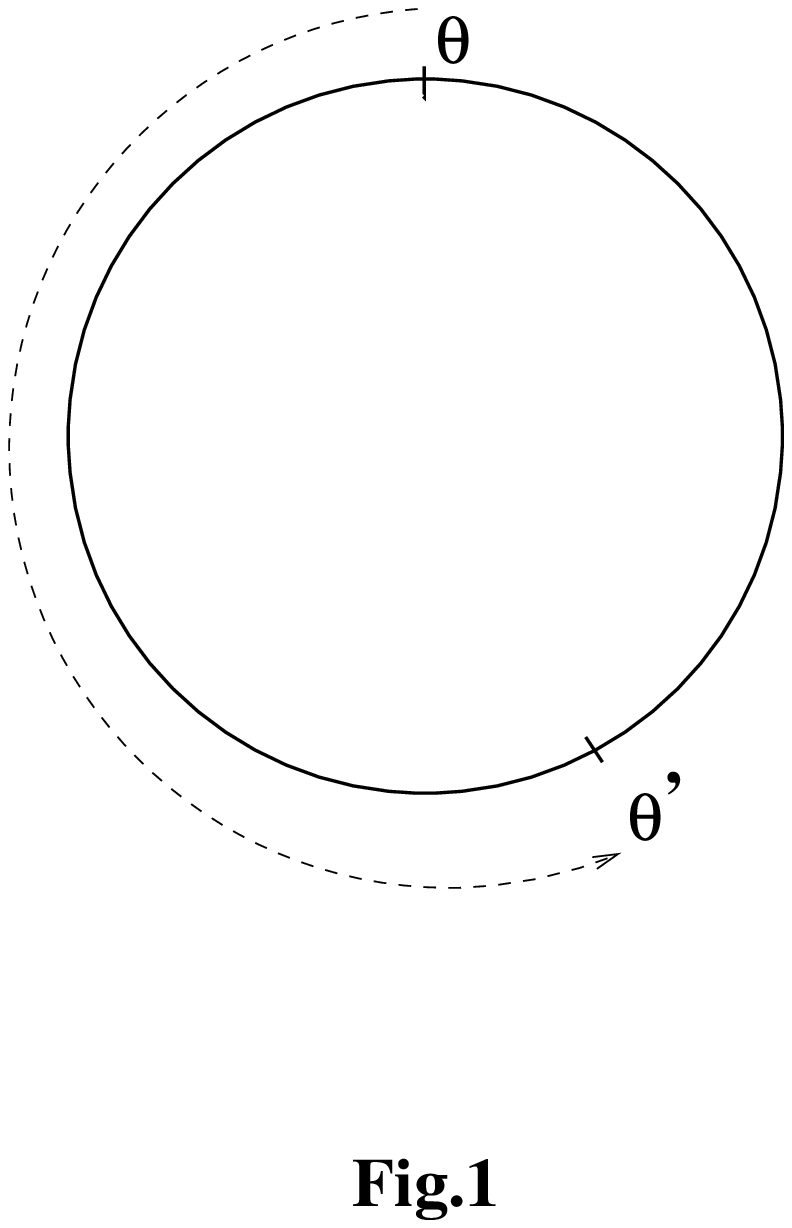}}
\end{center}
\vskip -0.5cm
\caption{}{Due to net positive contribution of the chemical potential
term to the free energy for clockwise variation of $\theta$, here
anti-clockwise variation from $\theta$ to $\theta^\prime$ may be
preferred compared to the shorter clock-wise variation.}
\label{Fig.1}
\end{figure}

\begin{figure}[h]
\begin{center}
\leavevmode
\epsfysize=20truecm \vbox{\epsfbox{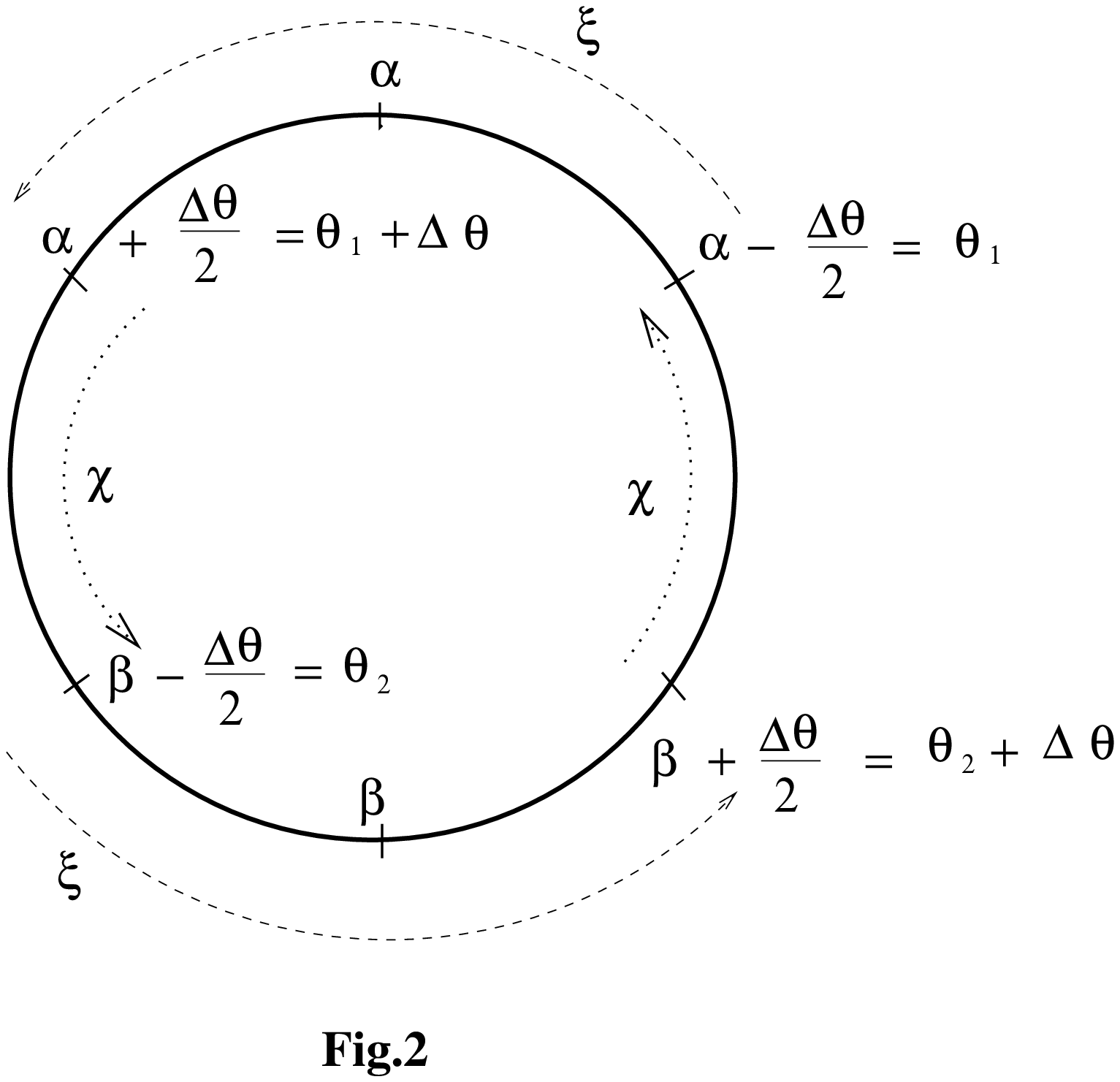}}
\end{center}
\vskip -0.5cm
\caption{}{Physical space $S^1$ consisting of two domains. Domains are of 
size $\xi$ and are shown by the dashed curves and inter-domain regions 
(of size $\chi$) are shown by the dotted curves. Arrows show anti-clockwise 
path taken in the physical space to calculate the winding of $\theta$.
The two random values of $\theta$ at the 
centers of the two domains are $\alpha$ and $\beta$. The anti-clockwise 
variation of $\theta$ as given by Eq.(7) within each domain is shown in the
figure. For notational simplicity, the angles at the two ends of the first
domain are denoted as $\theta_1 (\equiv \alpha - {\Delta \theta \over 2})$
and $\theta_1 + \Delta \theta (= \alpha + {\Delta \theta \over 2})$.
Similarly $\theta$ at the two ends of the other domain are denoted
as $\theta_2$ and $\theta_2 + \Delta \theta$.}
\label{Fig.2}
\end{figure}

\newpage

\begin{figure}[h]
\begin{center}
\leavevmode
\epsfysize=10truecm \vbox{\epsfbox{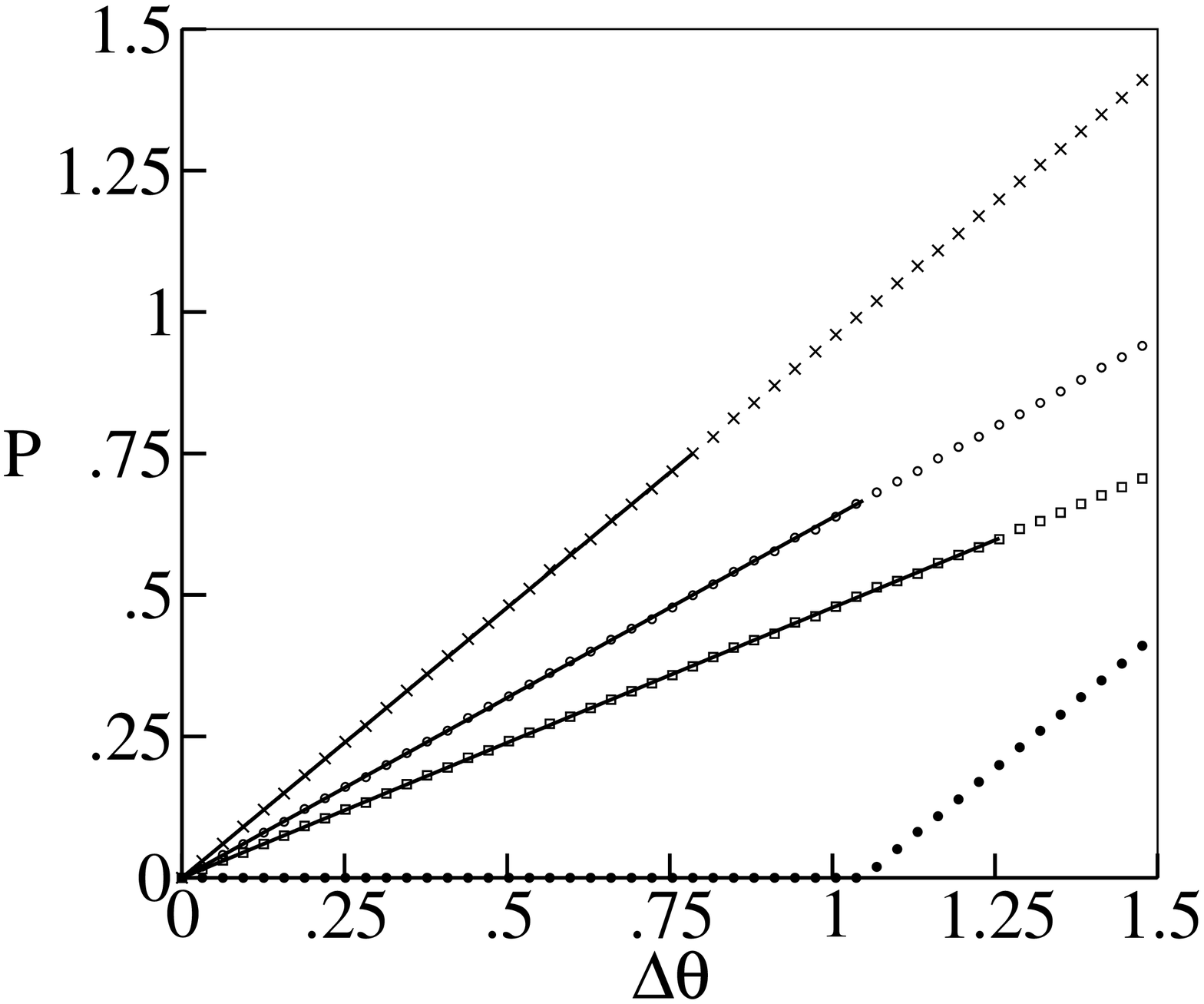}}
\end{center}
\vskip -0.5cm
\caption{}{Plot of the probability of Skyrmion formation (average Skyrmion
number) vs. $\Delta \theta$ with two domains. Probability 
of anti-Skyrmions remains zero. Symbols show the results of numerical
simulations while the solid line plots show the analytical result
from Eq.(22) (plotted only for $\Delta \theta < {\pi \over 
(\frac{\chi}{\xi} + 2)}$). Crosses, open circles, and open squares represent
the cases with ($\xi,\chi$) = (1,2), (1,1), and (2,1) respectively.
Solid dots show results for winding number 2 Skyrmions for the case
with $(\xi,\chi) = (1,2)$.}
\label{Fig.3}
\end{figure}

\newpage

\begin{figure}[h]
\begin{center}
\leavevmode
\epsfysize=8truecm \vbox{\epsfbox{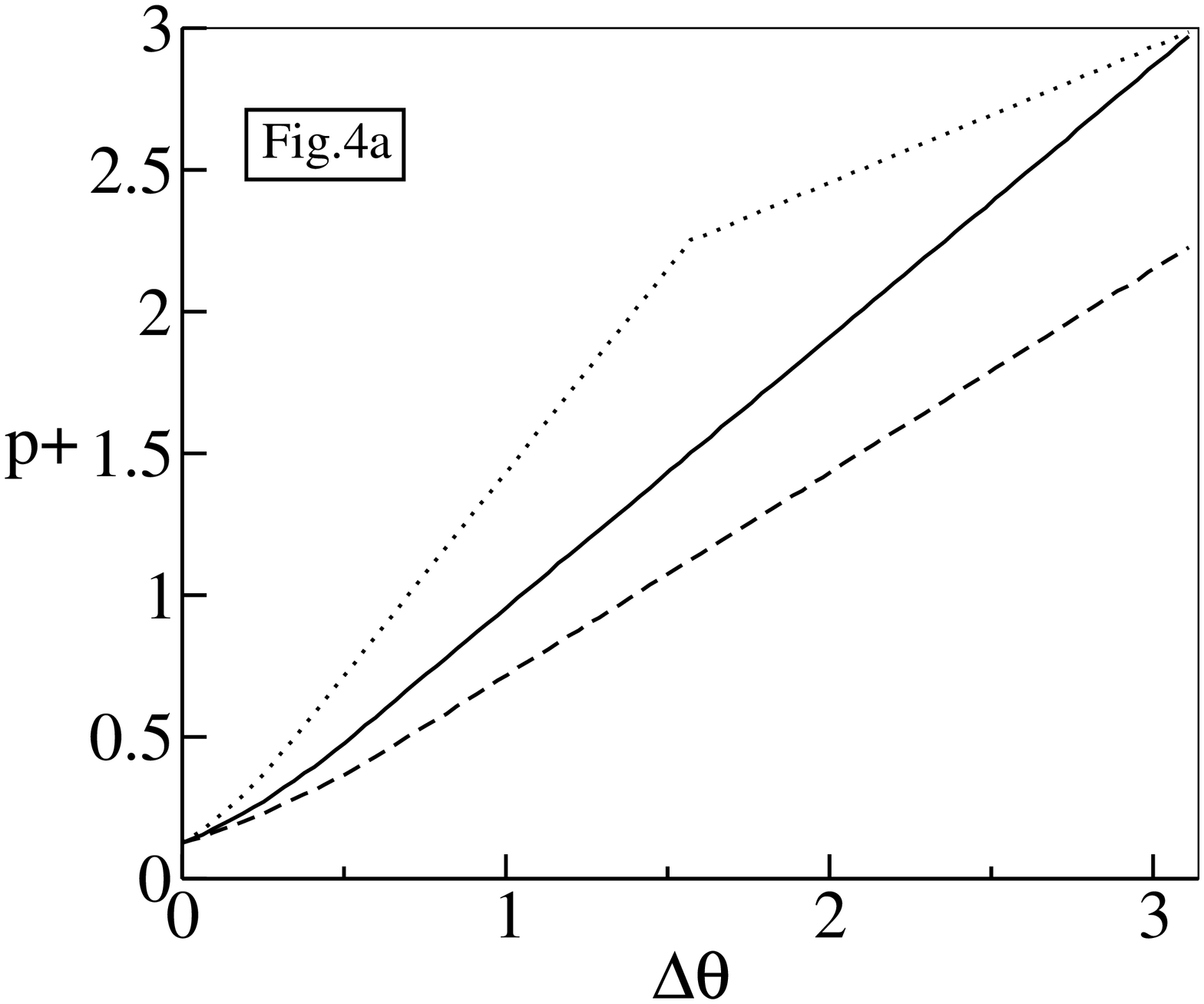}}
\epsfysize=8truecm \vbox{\epsfbox{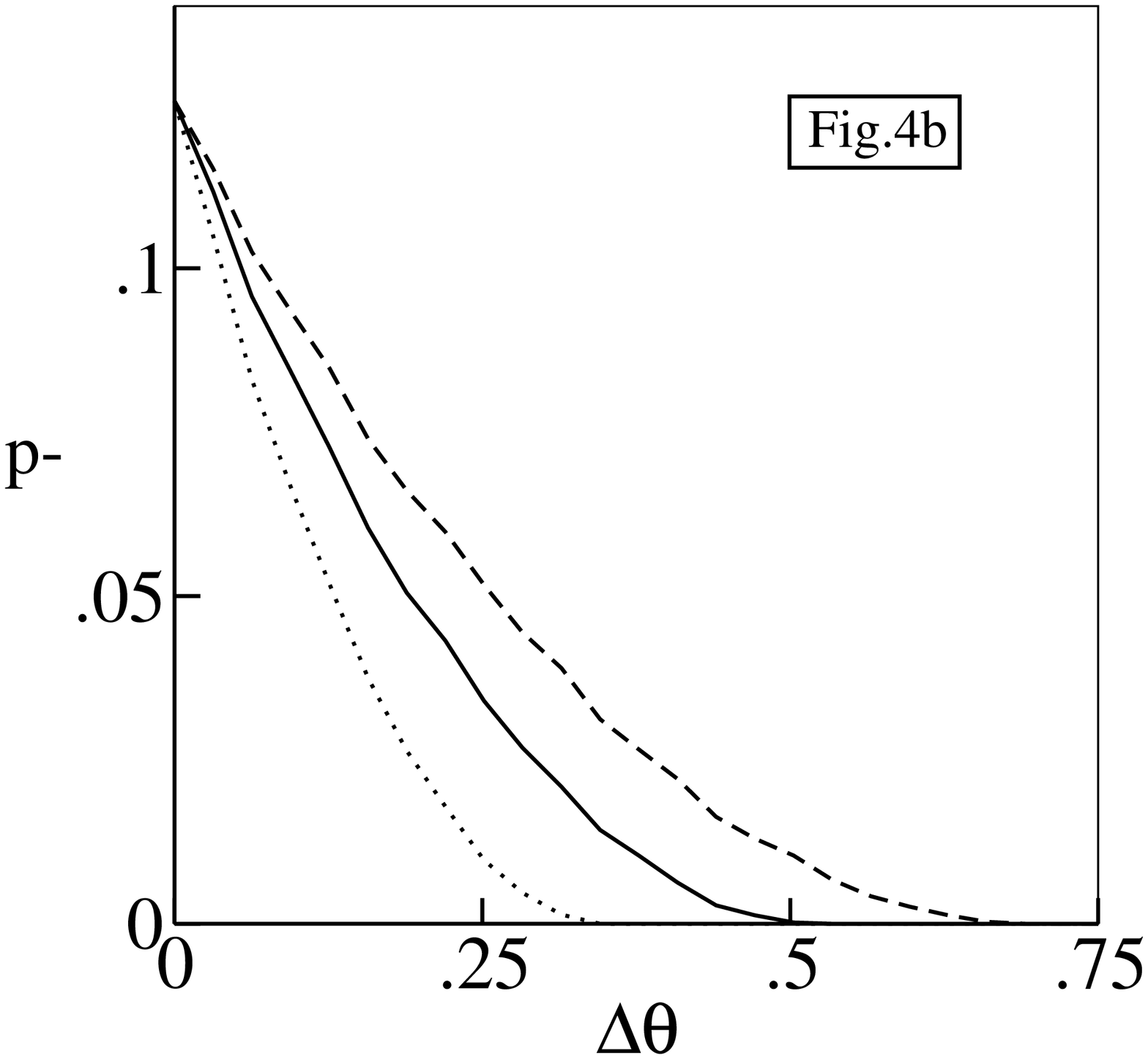}}
\epsfysize=8truecm \vbox{\epsfbox{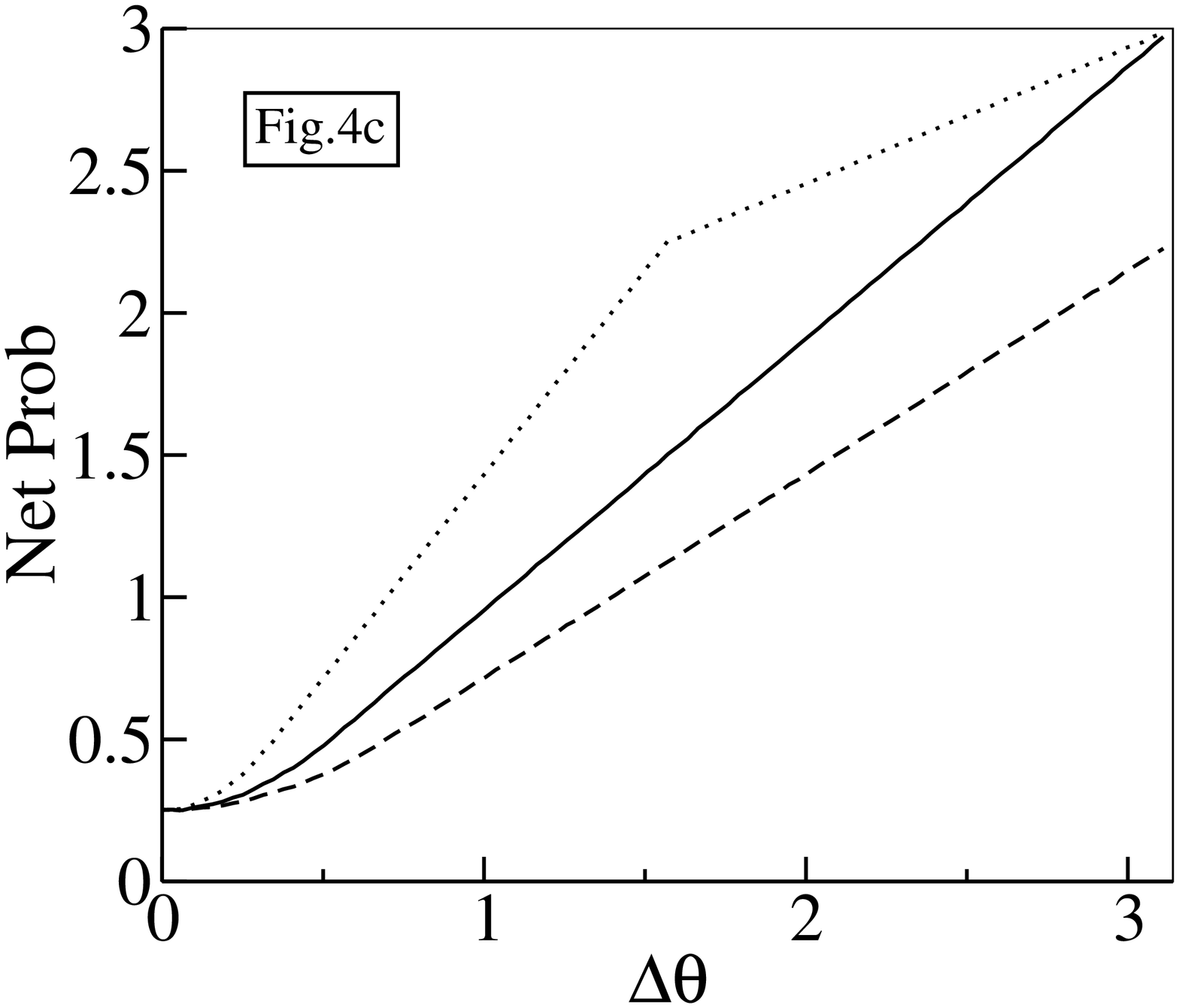}}
\end{center}
\caption{}{Dotted, solid,  and dashed plots show the probabilities
of Skyrmion formation (average Skyrmion number) for the choices
$(\xi = 1, \chi = 2), (\xi = 1, \chi = 1),$, and $(\xi = 2, \chi = 1)$
respectively. (a) shows the plots for the total probability (i.e., 
the average numbers) of positive winding Skyrmions. (b) shows plot
of anti-windings. (c) shows the sum of (a) and (b) (neglecting the sign of
the windings) giving the net average number of Skyrmions and antiSkyrmions.} 
\label{Fig.4}
\end{figure}

\newpage

\begin{figure}[h]
\begin{center}
\leavevmode
\epsfysize=8truecm \vbox{\epsfbox{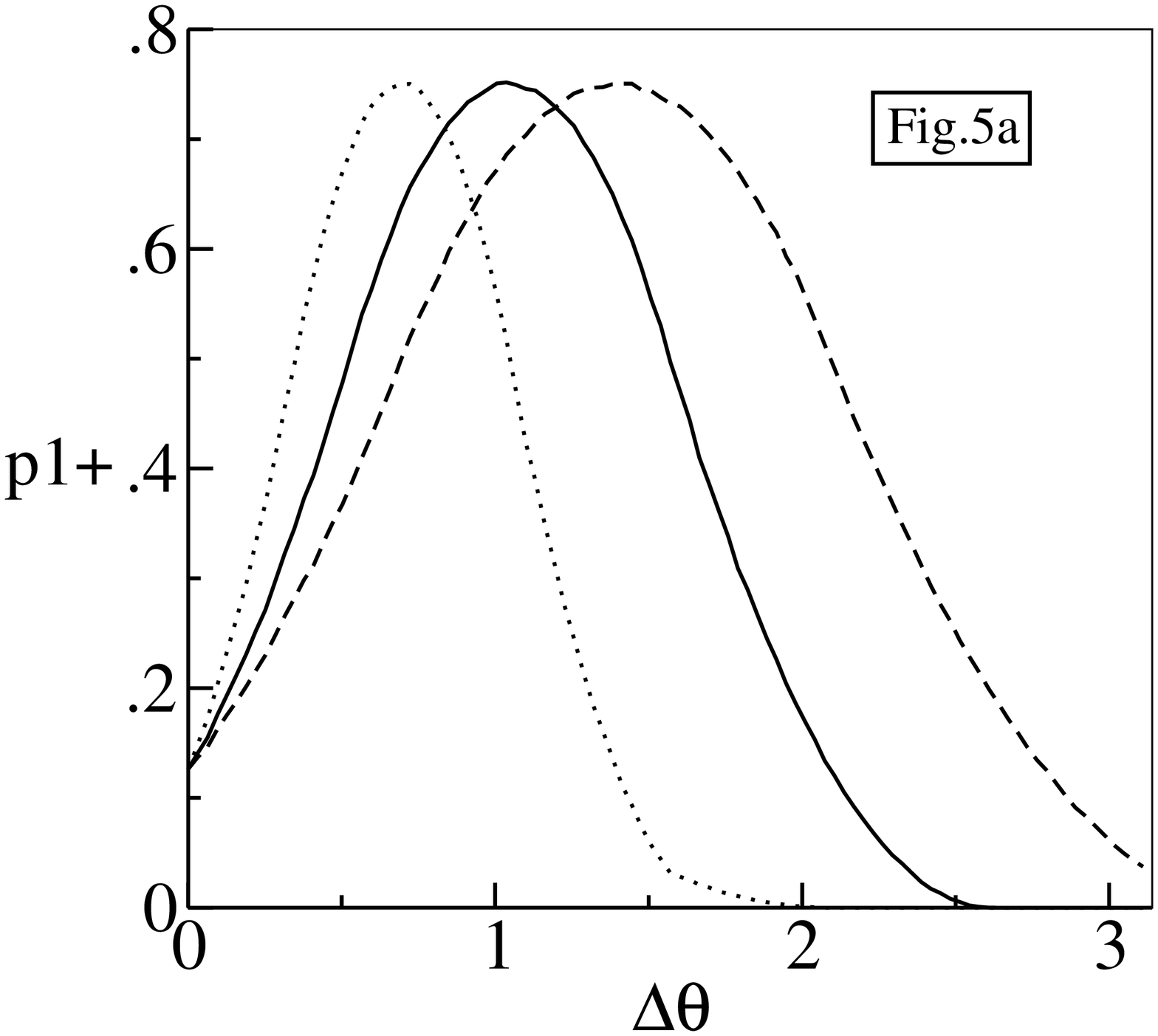}}
\epsfysize=8truecm \vbox{\epsfbox{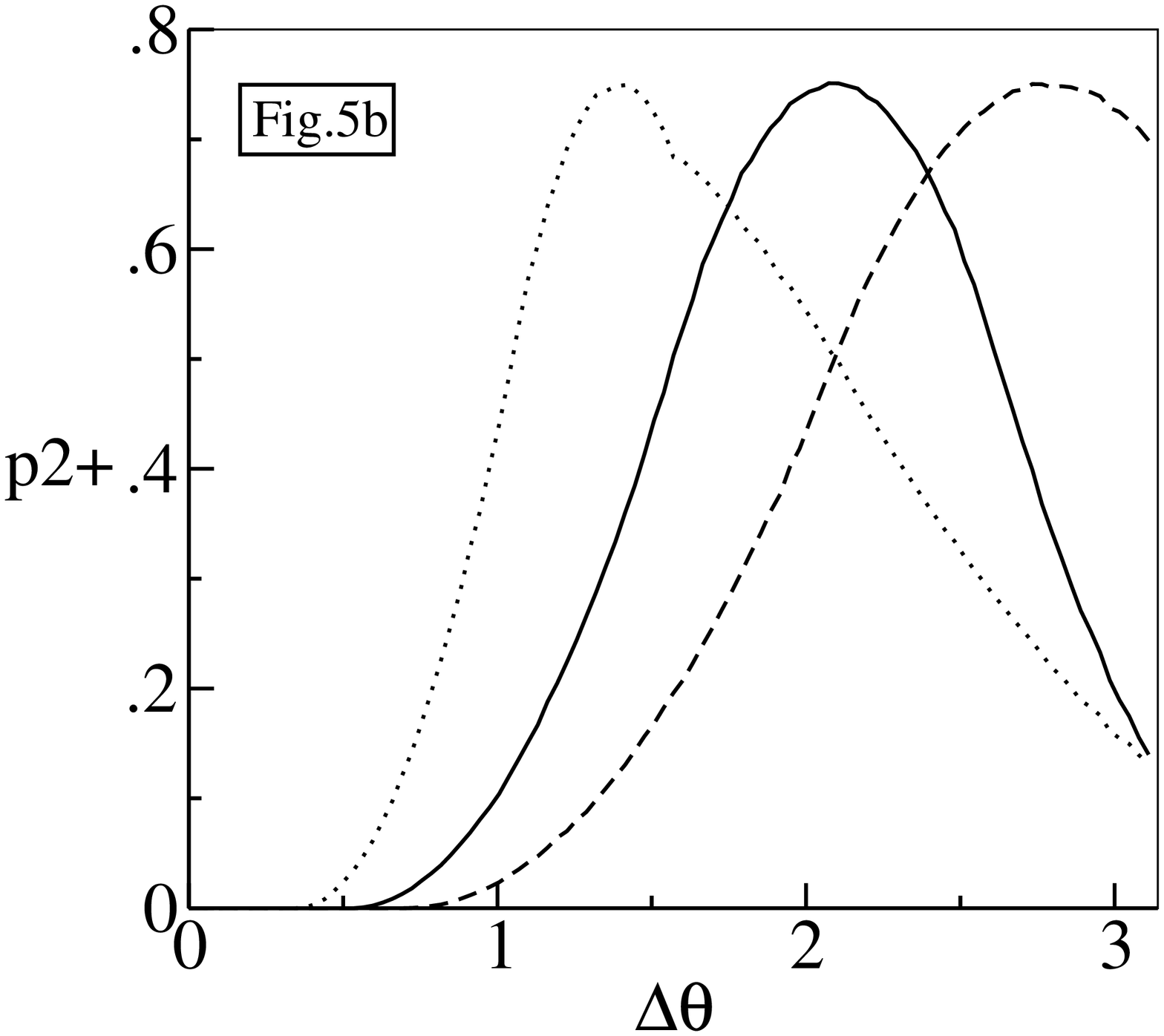}}
\epsfysize=8truecm \vbox{\epsfbox{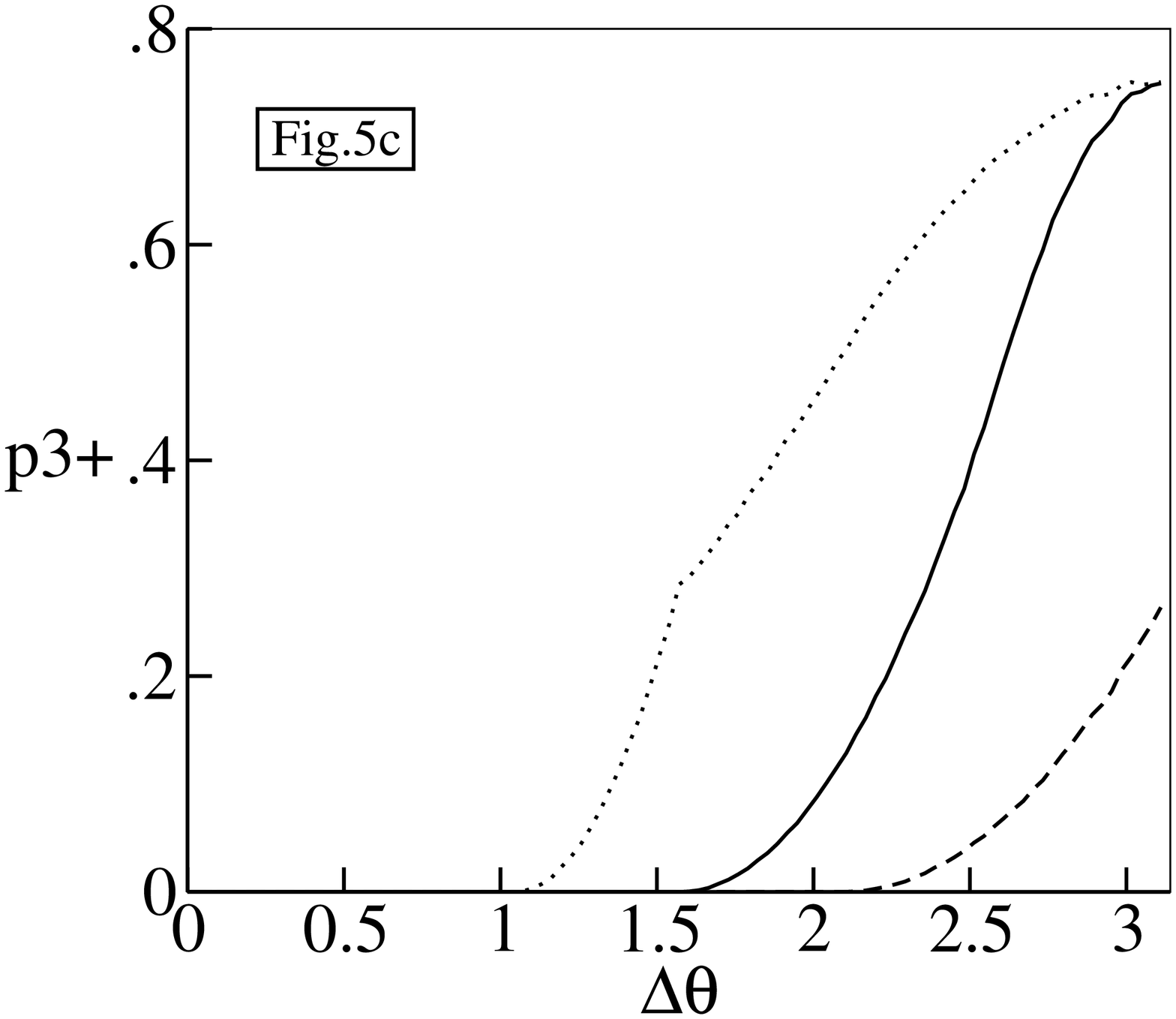}}
\epsfysize=8truecm \vbox{\epsfbox{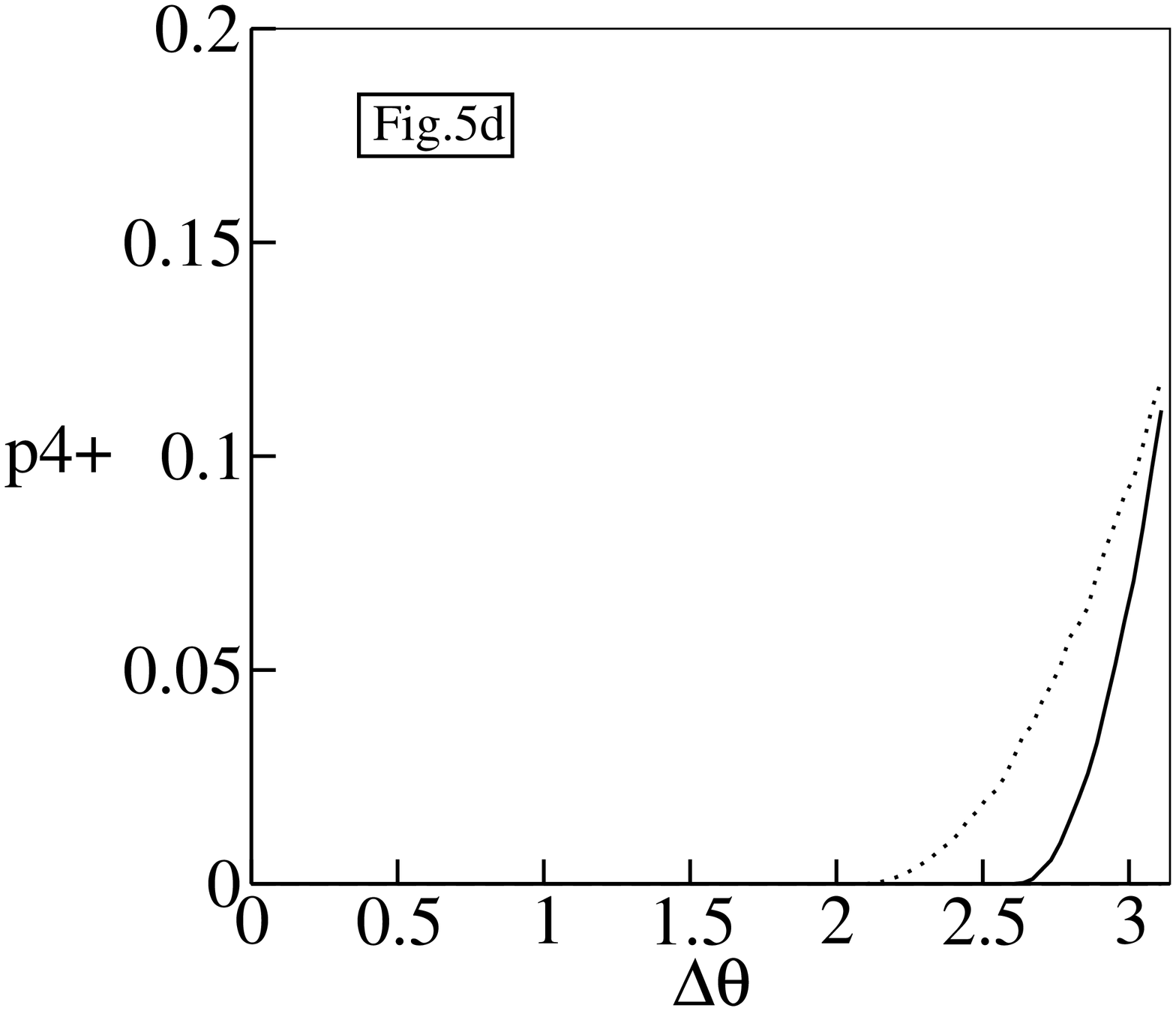}}
\end{center}
\caption{}{Plots are as shown in Fig.4. (a)-(d) give plots for
positive winding Skyrmions with winding 1,2,3, and 4, respectively.}
\label{Fig.5}
\end{figure}

\end{document}